\documentclass[twocolumn,showpacs,preprintnumbers,amsmath,amssymb]{revtex4-1}
\usepackage{graphicx}
\begin{document}
\def\half{\frac{1}{2}}
\def\d#1{#1^{\dagger}}
\def\ep{\epsilon}
\def\avg#1{\langle #1 \rangle}
\def\v#1{\mathbf{#1}}
\def\q{\v{q}}
\def\r{\v{r}}
\def\n{\v{n}}
\def\t{\mathcal{T}}
\def\ket#1{|#1\rangle}
\def\colvec#1{\left( \begin{array}{c} #1 \end{array} \right)}
\title{Localised zero-energy modes in the Kitaev model with vacancy-disorder}
\author{Santhosh G.}
\affiliation{Department of Physics, Indian Institute of Technology
Madras, Chennai 600036, India}
\author{V.\ Sreenath} 
\affiliation{Department of Physics, Indian Institute of Technology
Madras, Chennai 600036, India}
\author{Arul Lakshminarayan}
\affiliation{Department of Physics, Indian Institute of Technology
Madras, Chennai 600036, India}
\author{Rajesh Narayanan}
\affiliation{Department of Physics, Indian Institute of Technology
Madras, Chennai 600036, India}

\begin{abstract}
We study the effects of vacancy disorder on the Kitaev model defined on a hexagonal 
lattice.  We show that the vacancy disorder induces a zero-mode that is localized at the defect site. We derive analytical forms for these localized wave functions in both the gapped and gapless phases of the Kitaev model. We conjecture that the vacancy disorder can be utilized as a probe of the quantum phase transition (from the gapped to gapless phases) in this model. The behavior of the Inverse Participation Ratio (IPR) in the gapless phase and across the transition is also studied numerically. Comments are made about the behavior of site-site entanglement in the single particle states for the case of a single vacancy.    
\end{abstract}
\maketitle
\section{Introduction}
\label{Intro}
The effect of quenched disorder typified by impurities, lattice imperfections, and vacancies on 
condensed matter systems has been a source of intense scientific investigation in the recent past. 
In fact, the study of such ``frozen-in" disorder has led to the unravelling of a host of very interesting phenomena like infinite randomness fixed points~\cite{Fisher95}, quantum Griffiths effects~\cite{thill-huse-physa95,rieger-young-prb96}, and maybe even smearing of phase transition~\cite{vojta-prl03}. 
Even though as highlighted above multi-impurity effects can be extremely interesting, the study of 
a single impurity embedded in a host can also act as an efficient probe of the physical characteristics of the underlying bulk material, a situation exemplified by the case of local impurity acting as a probe of the 
order-paramater symmetry in unconventional superconductors~\cite{alloul-rmp}. This manuscript for the 
most part belongs to the latter genre wherein we study the role played by a single impurity in identifying the quantum phase transition inherent in the Kitaev model. 
   
   The Kitaev model has become one of the paradigmatic models that has been studied in various contexts ranging from strong correlation physics to topological quantum computation.  Its theoretical appeal lies in the fact that it  represents one of the few spin systems that can be solved exactly. The solution to the clean Kitaev model is effected by recasting the spin Hamiltonian into that of an equivalent Majorana hopping problem in the background of static $Z_2$ gauge configurations.  The exact solution of the model reveals both a gapless and gapped spin-liquid phases with a zero-temperature quantum 
  transition interpolating between these two phases. The gapless spin liquid phase is quite unique as it supports a spin-spin correlation function that is short ranged \cite{baskaran}, thus setting it apart from other spin-liquid phases studied so far. It also supports fractionally charged topological excitations both Abelian and non-Abelian that can be plausibly used to perform quantum computations.   
   Apart from its utility for topological quantum computation or for its usefulness in studying spin-liquid 
   ground states, the spin-1/2 Kitaev model defined on a two dimensional hexagonal lattice has become a powerful test-bed example to study various fundamental concepts in the field of strong correlation physics.  For instance,  it has been used to study fractionally charged excitations that occur in topological insulators thereby providing a beautiful higher dimensional extension \cite{dunghailee} of the Jackiw-Rebbi theory \cite{jackiw}, \cite{su-schrieffer-heeger}, that describes charge fractionalization in one dimensional system.  It has also been utilized to study dynamics of quantum quenches across the critical region \cite{krishnendu}. Moreover, there now exists higher dimensional realizations of the Kitaev model \cite{naveen}, and also extension to higher values of spin \cite{shankar-diptiman}
    
  As is clear from the preceding paragraph the clean Kitaev model has been the subject of intense 
  scientific investigations in the last few years. However, apart from a few notable exceptions, (detailed in the next paragraph),  one area that has been rather neglected is the study of the effect of impurities on the Kitaev model. This is a particularly glaring deficiency as now there exists proposals for experimental  
  realizations of the Kitaev model \cite{jackeli}, \cite{jackeli2}, \cite{duan}. Thus, the study of impurity effects gain an added significance as realistic systems are seldom clean.  
  
  Now, the simplest form of quenched disorder involves studying the effect of a single impurity on the 
  bulk system. Such a study was undertaken by \cite{kusum} for the case of a single magnetic impurity that was embedded in the host Kitaev model. It was shown that coupling of an impurity to the host Kitaev system leads to an unusual Kondo effect that is sensitive to the topological transition in the Kitaev model. 
  In a related work, Willans et. al \cite{willans} showed that disorder in the form of a single vacancy binds a flux which in turn gives rise to a local moment. Furthermore they showed that this moment leads to a vacancy susceptibility that diverges logarithmically as a function of the applied field, (for weak applied fields). 
  
  While it is true that impact of quenched disorder has received very scant attention, the effect of impurities on allied models have been rather well studied. More specifically: The Kitaev model 
  maps onto a fermionic model displaying bi-partite hopping on a hexagonal lattice in the 
  background of $Z_2$ gauge fields. Now, the impact of quenched disorder on similar bi-partite hopping models have been studied in the context of Anderson localization. In a seminal work, (see \cite{gade} \cite{gadewegner}), it was shown that 
the quenched impurities lead to a divergent Density of States, (DoS) in models that 
display bi-partite hopping. More specifically, by using a field theoretical formulation, 
these authors showed that a random-mass form of disorder, 
(in addition to a random vector potential) would lead to a highly divergent DoS that 
conforms to the functional form, $\rho(E) \sim \frac{1}{E}e^{-|\ln E|^{1/x}}$, with $x=2$. This 
faster than power-law divergence of the DoS should be contrasted with the results of 
Ludwig et. al \cite{ludwigetal} wherein they studied a random vector potential model 
with bi-partite hopping. These authors showed that the DoS in these models diverge as a 
power-law $\rho(E) \sim E^{-1+2/z}$, where $z$ is a continuously varying dynamical exponent. 
Now, the field theoretical treatment of Gade and Wegner suffered from a slight draw-back: It did not 
provide a physical frame-work wherein the diverging DoS at the band center could be understood. 
This situation was remedied in the work of Motrunich et. al \cite{motrunichet.al} wherein an intuitive 
and physically appealing argument was provided for the origin of the diverging DoS and the low-lying 
states that was causing it. They further argued that the DoS indeed does diverge with a functional form 
analogous to that derived by Gade and Wegner, however, with the exponent $x=3/2$ instead of $2$ 
obtained in \cite{gadewegner}. Another extremely relevant physical context 
 wherein such disordered bi-partite hopping models holds relevance is provided by the case of graphene. In graphene, due to open surfaces and substrates, disorder is an ever present bug-bear.  One form of disorder that has been relatively well studied in the context of graphene is vacancy disorder. This type of disorder arise naturally in the case of irradiated graphene wherein the carbon atoms are knocked out of graphene planes.  The impact of such vacancies on the electronic properties of graphene were investigated in a series of papers \cite{CastroNeto},  \cite{CastroNetoPRL}. 
 The influence of various other forms of disorder, (also inclusive of the case of vacancy disorder) 
 on the electronic properties of graphene was studied in \cite{Mirlin1}, \cite{Mirlin2}.
  
  In this paper we will study the influence of vacancy disorder in the Kitaev model. More specifically, we focus our attention mainly on the structure of the wavefunction at the site of a vacancy disorder. 
  We show that  vacancy disorder gives rise to a ``zero-mode" that is localized at the impurity site. 
   As we shall see later on in this paper, this zero-mode exists in both the gapped and gapless phases of this model and is a consequence of the particle-hole symmetry of the bi-partite hopping problem. However, as will be shown in the bulk of this mansuscript, the functional form of the zero-mode is quite different in two phases thereby providing an invaluable tool for distinguishing different phases of this model. The interesting question of impact of many vacancies on the Kitaev model will be addressed in a
   future publication \cite{us_future}. 
   
   The paper is organized in the following manner: In Sec.~\ref{model} we recapitulate the 
mapping of the Kitaev model into a non-interacting Majorana fermion problem by following the Jordan-Wigner 
fermionization scheme. Sec.~\ref{single vacancy} is devoted to the analytic derivation of the zero-mode wave function that is localized at the vacancy site. The functional form for the 
wave function is established for both the gapless phase,~\ref{Gapless} and for the 
gapped spin-liquid phases~\ref{Gapped}. In ~\ref{IPR} we will delve into the issue of 
the ``flow" of the Inverse Participation Ratio (IPR) of the localized wavefunction as a function of the coupling parameters in the Kitaev model. In the same subsection we will also touch upon related entanglement measures. Finally, we will end with a concluding section, Sec.~\ref{Conclusion} wherein our results will be briefly reviewed and placed in context of the existing literature in this field. We will also briefly mention some of the open problems that remain to be tackled in this subsection.

\section{Kitaev Model}
\label{model}
\begin{figure}[t]
\begin{center}
\includegraphics[width=2.5in]{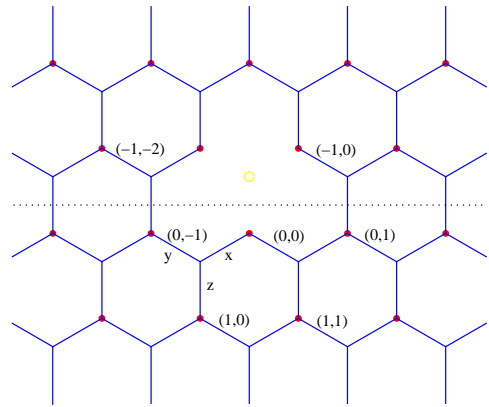}
\caption{Hexagonal lattice with a vacancy.  A unit cell contains two points, one point marked red and another unmarked point forming a bond.  The unit cells are labelled  using pair of integers $(j,l)$ and the convention used is made clear through explicit labeling of some unit cells. The vacancy is denoted by a yellow circle. The dotted line separate the part  of the lattice with $j\geq0$ from the part with $j<0$.}
\label{figLattice}
\end{center}
\end{figure}
 In this section we give a short introduction to the model and its properties for later reference. It also serves to set the notations for the rest of this paper.
The model comprises of spins 
residing on the sites of a honeycomb lattice as shown in Fig.~\ref{figLattice}. 
The spins interact with each other via nearest neighbour 
coupling which is dependent on the bond orientation. These orientations are labeled as  
 as $x$, $y$ and $z$ in the figure, Fig.~\ref{figLattice}. Also, as represented in Fig.~\ref{figLattice},  
 the $y$-link is taken to be the basis, with a two ``atom" unit cell: The red colored lattice point
denoting the A-sublattice and the uncolored point indicative of the B-sublattice respectively. They 
are connected to each other via the $y$-link.  A point in the lattice is thus labeled by a triplet of numbers $(j,l,\mu)$ where $j,l$ denote the unit cell and $\mu=1(2)$ correspond to the A(B) sublattice. Thus, under this labeling scheme the Hamiltonian is expressed as:
\begin{eqnarray}
 H&=&\sum_{j,l} \big [ J_x \sigma_{j,l,1}^x \sigma_{j,l-1,2}^x +J_y \sigma_{j,l,1}^y \sigma_{j,l,2}^y\nonumber\\
  &&+ J_z \sigma_{j,l,1}^z \sigma_{j-1,l-1,2}^z \big ].
\end{eqnarray}
Here, as usual the $\sigma $s are the usual Pauli matrices that  represent the spin variables.

As briefly discussed in the introduction, Sec.~\ref{Intro}, this spin-model can be mapped onto a 
Majorana Fermion hopping problem. Different methods can be adopted to effect this transformation. In this manuscript we shall employ the Jordan-Wigner fermionization scheme as employed by Feng et. al., 
\cite{Feng} in this context. Define the Jordan-Wigner tail operator  as
\begin{eqnarray}
 K(j,l,\mu)= \prod_{ (j,l,\mu)> (m, n,\nu) } \sigma_{m,n,\nu}^z ,
\end{eqnarray}
where $(j,l,\mu) >(m,n,\nu)$ if  $j>m$ or $j=m,l>n$  or $j=m,l=n,\mu>\nu$. Now the Majorana fermion operators can be defined  as:
\begin{equation}
 \psi^a_{j,l,\mu}=K(j,l,\mu) \sigma_{j,l,\mu}^x, \quad \psi^b_{j,l,\mu}=K(j,l,\mu) \sigma_{j,l,\mu}^y.
\end{equation}
In terms of these operators the Hamiltonian takes the form
\begin{eqnarray}
 H&=&i\sum_{j,l} \big[  J_x \psi^a_{j,l,1} \psi^b_{j,l-1,2}+ J_y \psi^a_{j,l,1} \psi^b_{j,l,2}\nonumber\\
   && \quad + J_z D_{j,l} \psi^a_{j,l,1} \psi^b_{j-1,l-1,2}
\big].
\end{eqnarray}
Here,  the operators $D_{jl}=i \psi^b_{j,l,1} \psi^a_{j-1,l-1,2}$, defined on the $z$-links, 
is Hermitian and  commutes among themselves and with the Hamiltonian reflecting the local symmetry of the Kitaev model. It can be shown that the operators  $D_{jl}$ have eigenvalues $\pm 1$.  The Hamiltonian gets block diagonalised into different sectors corresponding to different sets of eigenvalues of $D_{jl}$. In each of these sectors, the Hamiltonian becomes a quadratic fermionic system obtained by replacing each $D_{jl}$ by its eigenvalue and can be re-cast into the form:
\begin{eqnarray}\label{HamiltonianStructure}
 H=\frac{1}{4} \psi^T iA \psi,
\end{eqnarray}
where $\psi\equiv (..,\psi^a_{j,l,1} \psi^b_{j,l,2},..)^T$ and $A$ is an antisymmetric matrix. 

Thus, as alluded to in the introduction the Kitaev model has been mapped onto a non-interacting Majorana fermion problem in the background of static $Z_2$ gauge field.

If there are $N$ number of unit cells, we have $2N$ spins and the Hilbert space is $2^{2N}$ dimensional. Since there are $N$ $z$-links, there are $2^N$ sectors each of which has dimension $2^N$ corresponding to $2N$ Majorana fermions.

For further calculations, let us first see how the eigenvectors/values of the coefficient matrix $iA$ are related to the fermionic excitation modes of the system. Since $A$ is antisymmetric, the eigenvalues of $iA$ comes in pairs $-\ep_i,\ep_i$ with eigenvectors $v_i,v^*_i$ respectively, where $\ep_i \geq 0$. We can choose the eigenvectors to be orthonormal since $iA$ is Hermitian. 
Define fermion operators $d_i=\frac{1}{\sqrt{2}}  \psi^{T} v_i $.  It is easily seen that these operators obey $\{\d{d}_i,d_j\}=\delta_{i,j}$.
We get,
\begin{eqnarray}\label{HamiltonianGeneral}
 H=\sum_{i=1}^{N} \ep_i \left(\d{d}_i d_i-\frac{1}{2} \right).
\end{eqnarray}

It is known that the ground state of the Hamiltonian lies in a sector wherein all the  $D_{jl}$ operators take the eigenvalue $+1$~\cite{Kitaev, Lieb}. By making use of the translational symmetry in the problem, a solution of the model can be effected by going into Fourier transformed representation . Thus, the Hamiltonian re-expressed in terms of the Fourier transformed variables,
  $(\psi^{a}_{\v{k}}\;\psi^{b}_{\v{k}})$ $=\sum_{j,l}e^{-i\v{k} \cdot \v{r}_{j,l}} (\psi^{a}_{j,l,1} \; \psi^{b}_{j,l,2})/\sqrt{2N} $, (where $\v{r}_{j,l}=j \v{n}_1+l\v{n}_2$) reads, 
\begin{eqnarray}
 H=\frac{1}{2} \sum_{\v{k}} (\psi^a_{-\v{k}}\;\psi^b_{-\v{k}}) 
\left(\begin{array}{cc}
 0& i\phi(\v{k})\\
-i\phi^*(\v{k})&0
 \end{array}
 \right)
\left(
\begin{array}{c}
 \psi^a_{\v{k}}\\ \psi^b_{\v{k}}
\end{array}
\right).
\end{eqnarray}
Here, $\phi(\v{k})= 2(J_x e^{-ik_2}+J_y+J_z e^{-i(k_1+k_2)})$ and $k_i=\v{k} \cdot \v{n}_i$. 
The eigenvalues are $\pm |\phi(\v{k})|$ and the fermionic excitations are given by
\begin{eqnarray}
H=\sum_{\v{k}} |\phi(\v{k})| \left( \d{d}(\v{k}) d(\v{k})-\frac{1}{2} \right),
\end{eqnarray}
where $d(\v{k})=\frac{1}{\sqrt{2}}[ \psi^a_{\v{k}}+ i \frac{\phi(\v{k})}{|\phi(\v{k})|} \psi^b_{\v{k}} ]$. The excitation spectrum is gapless if there exist points where $|\phi(\v{k})|=0$ which is possible only if following condition is satisfied:
\begin{equation}\label{gapCondition}
(J_x-J_y)^2<J_z^2<(J_x+J_y)^2.
\end{equation}
  The gapless phase is characterised by Fermi points where the $\phi(\v{k})$ vanishes. There is a quantum phase transition from the gapless- to the gapped-phase as the parameters cross the conditions in Eq.~\ref{gapCondition}. This quantum phase transition is the one that we wish  to probe via a single vacancy disorder.

\section{Single vacancy}
\label{single vacancy}
In this section, we study the nature of the wave-function at the site of a single vacancy. The analytic 
functional form is derived for both the gapped and gapless phases. 
 
For the sake of concreteness consider the Kitaev model with a vacancy at the B-site in the unit-cell $(-1,-1)$, (see Fig.~\ref{figLattice}). Note that the Jordan-Wigner construction goes through with 
the tail operator $K(j,l,\mu)$ missing $\sigma_{-1,-1,2}^z$ for all $(j,l,\mu) > (-1,-1,2)$. As the system is no longer translationally invariant one cannot use Fourier transform to solve the problem. However, the general structure of the Hamiltonian, Eq.~\ref{HamiltonianStructure}, remains with $A$ now being a $(2N-1) \times (2N-1)$ matrix obtained by removing the row and column corresponding to the site $(-1,-1,2)$ from the matrix $A$ in Eq.~\ref{HamiltonianStructure}. 
 Thus, we have $N-1$ eigenvectors forming pairs as described in Sec.~\ref{model} and one unpaired eigenvector  denoted by $\tilde{v}$. This eigenvector should be real with zero eigenvalue because of the $\ep \leftrightarrow -\ep$ symmetry briefly alluded to in Sec.~\ref{model}. The $N-1$ pairs can be combined to form $N-1$ complex fermion operators $d_i$ leaving behind a single unpaired mode. This unpaired eigenvector forms a Majorana mode, $\tilde{d}=\psi^T \tilde{v}$ as $\d{\tilde{d}}=\tilde{d}$ and $\tilde{d}^2=\mathbf{1}$. Note that by removing a spin at $(-1,-1,2)$, we have left out one other Majorana fermion operator from the Hamiltonian; $\psi^b_{0,0,1}$ which would have formed the operator $D_{0,0}$ with $\psi^a_{-1,-1,2}$ had the spin been present. Thus a complex fermion mode can be constructed from these two Majoran modes as $d_N=\frac{1}{2}( \tilde{d}+i \psi^b_{0,0,1} )$ which will be a zero-energy excitation of the Hamiltonian which, again, has the same form as Eq.~\ref{HamiltonianGeneral}.  However, since $D_{00}$ is not present in the Hamiltonian, the number of  $D_{jl}$ operators is now $N-1$. Thus, now there are $2^{N-1}$ sectors each with $N$ complex fermions. Hence as expected the total dimension of the system is $2^{2N-1}$.

Now that the above discussion has clearly established that a single vacancy induces a  zero-energy fermionic excitation mode in the Kitaev model, let us turn our attention to the analytic structure of these modes. To do so, we employ a method developed by Pereira et. al \cite{CastroNeto}, \cite{CastroNetoPRL} in the context of zero modes arising out of a vacancy defect in graphene.  
This adaptability of the technique developed for the case of graphene to the Kitaev model is not so surprising as they both give rise to similar fermion hopping problems. Unlike the case of graphene studies by Pereira et. al. \cite{CastroNeto}, \cite{CastroNetoPRL}, where one is restricted to the isotropic case  $J_x=J_y=J_z$, here we consider the general anisotropic hopping problem and obtain expressions for the zero mode for the parameter regimes corresponding to both gapped- and gapless-phases of the clean model. More specifically, we obtain an asymptotic form for the defect wave-function 
in the gapless phase, whereas one can evaluate an exact form of the wave function in the gapped phase.

Before obtaining explicit expression for the zero modes of $iA$ with B-site vacancy, let us first see how they are related to the corresponding zero mode when the vacancy site is in the A-sublattice. Let us introduce the notation $\v{r} \equiv (j_r, l_r)$ and $A_{\mu,\nu}(\v{r},\v{r}')\equiv A_{(j_r,l_r,\mu),(j_{r'},l_{r'},\nu)}$. The clean model has the symmetry given by $\tau^x A(\v{r}+\v{\rho},\v{r}+\v{\rho'}) \tau^x=-A(\v{r}-\v{\rho},\v{r}-\v{\rho'})$ for any $\v{r}$, where $\tau^x$ is the  Pauli matrix. But the vacancy breaks the translational invariance. Let $V_{\mu,\v{r}_0}$ be the matrix to be added to $A$ to create the vacancy, by removing corresponding matrix elements from $A$, at position $\v{r}_0$ in the sublattice $\mu$. Now, $\tau^x V_{1,\v{r}_0}(\v{r}_0+\rho,\v{r}_0+\rho') \tau^x=-V_{2,\v{r}_0}(\v{r}_0-\v{\rho},\v{r}_0-\v{\rho}')$. If $\phi_{2,\v{r}_0}(\v{r}) \equiv \phi(\v{r}-\v{r}_{0})$ is an eigenvector of $i(A+V_{2,\v{r}_0})$ with eigenvalue $\lambda$, then it follows that  $\phi_{{1},\v{r}_{0}}(\v{r})\equiv \tau^x \phi(\v{r}_0-\v{r})$  is an eigenvector of $i(A+V_{1,\v{r}_0})$ with eigenvalue $-\lambda$. Thus we need to find the localised zero mode with one type of vacancy only, the other obtained from the relation given above. In the discussion hereafter, $\phi_{{\mu},\v{r}}$ always denote the zero mode created by $\mu$-sublattice vacancy at site $\v{r}$.

 Consider a B-site vacancy in the unit cell $(-1,-1)$ as shown in the Fig.~\ref{figLattice}. 
 The eigenvalue equation of $iA$ for the zero eigenvalue decouples the A- and B-sublattice amplitudes. Denoting the A-sublattice amplitude by $a_{jl}$, we get the corresponding eigenvalue equation as
\begin{equation}\label{eigenvalueEquation}
J_y a_{jl}+ J_x a_{j,l+1}+ J_z a_{j+1,l+1}=0.
\end{equation}
This equation hold true everywhere except for $j=-1=l$, where it no longer applies due to the vacancy. A similar equation can also be written down for the B-sublattice amplitudes. In that case, the  corresponding equation is satisfied by choice of them being equal to zero identically. 
Thus, we have $\phi_{2,(-1,-1)}(\v{r})=(a_{j_r l_r} \quad 0)^T$.

To solve Eq.~\ref{eigenvalueEquation}, following the procedure of Pereira et. al.~\cite{CastroNetoPRL},  the lattice is divided into two parts: $j \geq 0$,  and $j<0$, the  parts that lies below and above the dotted line respectively in Fig.~\ref{figLattice}. Eq.~\ref{eigenvalueEquation} is solved separately in these two regions and a boundary matching condition is imposed at the dotted line in Fig.~\ref{figLattice}. Applying periodic boundary condition along the horizontal direction, a Fourier transformation ,  $a_j(q) = \sum_l e^{-i q l} a_{jl}$, reduces Eq.~\ref{eigenvalueEquation} to 
\begin{eqnarray}\label{recursionFourier}
a_{j}(q)=-f(q) a_{j-1}(q),\quad f(q)=\frac{(J_y e^{-iq}+J_x)}{J_z}.
\end{eqnarray}
The solutions are given by
\begin{eqnarray}
a_{j}(q)= \left\{  \begin{array}{c}
	                  \left[ -f(q) \right ]^j a_{0}(q)\quad \forall\; j>0 ,  \\
 			\left[ -f(q) \right ]^{j+1} a_{-1}(q) \quad \forall \; j<-1.
	          \end{array} \right.
\end{eqnarray}
Seeking solutions that decay as a function of the distance from the vacancy site, we get the following conditions: $a_0(q)$ is non-zero only if $|f(q)|<1$, and $a_{-1}(q)$ is nonzero only if $|f(q)|>1$. Note that these conditions require that the $j<0$ and $j\geq0$ regions have complementary sets of wave-vectors contributing to the eigenvectors. The boundary condition at the interface is now implemented as
\begin{eqnarray}
\sum_{q} e^{iq(l+1)}\left[ 
a_0(q) +\frac{J_x + J_y e^{-iq}}{J_z} a_{-1}(q)
\right]=0,
\end{eqnarray}
except for $l=-1$.
This set of equations is satisfied by the choice $ a_0(q)=\Theta(1-|f(q)|)$ and $ \frac{(J_x + J_y e^{-iq})}{J_z}a_{-1}(q)=\Theta(|f(q)|-1)$. It is easily checked that the condition $|f(q)| \leq  1$ can be satisfied for parameter values that obey Eq.~\ref{gapCondition}. For other parameter values, corresponding to the gapped-phase of the clean model, $|f(q)|$ will either be less than $1$ or will be greater than $1$ for all values of $q$, thereby giving us trivial solution in one of the two regions. We now consider these two parameter regions separately. 
\subsection{Gapless phase}
\label{Gapless}
First, consider the gapless phase, where  we have a set of $q$ values in the range $(q^*, 2\pi-q^*)$ that satisfy $|f(q)|<1$, where $\cos(q^*)= \frac{J_z^2-J_x^2-J_y^2}{2J_x J_y}$ with $q^*\in(0,\pi)$.
Its complement in $[0,2\pi)$ gives the set of $q$ values contributing to eigenvector in the $j<0$ region. The eigenvector for the $j>0$ region is now constructed by taking the inverse Fourier transform of $a_j(q)$. Thus, we have
\begin{eqnarray}\label{eigenvectorGapless}
a_{jl}&\sim \Re \left\{ 
 \int_{q^*}^{\pi} dq\, e^{iq(l-j/2)} \left[  - \ep(q) e^{i\theta(q)} \right]^j
\right\}.
\end{eqnarray}
Here, $\ep(q)=[J_x^2+J_y^2+2J_x J_y \cos(q)]^{1/2}/J_z$ and $\tan(\theta(q))=(J_x-J_y)\tan(q/2)/(J_x+J_y)$. Notice that $\ep(q)$ decreases monotonically from its maximum value $1$ at $q^*$ to $|J_x-J_y|/J_z$ at $q=\pi$. For asymptotically large values of $j$, the dominant contribution comes from the region around $q^*$ . Therefore, expanding around $q^*$,  the above equation, Eq.~\ref{eigenvectorGapless} can be written in terms of its asymptotic form as
\begin{eqnarray}\label{asymptotic}
a(x,y)\sim \Re\left\{
 \frac{e^{iq^*x/\sqrt{3}+i2(\pi+\theta^*)y/3}}{\alpha y/\sqrt{3}-i 2\beta y/\sqrt{3} -ix}
\right\} .
\end{eqnarray}
The parameters $\alpha$, $\theta^*$, and $\beta$ are given by $\alpha=2J_x J_y \sin(q^*)/J_z^2$, $\theta^*=\theta(q^*), \beta=(J_x^2-J_y^2)/J_z^2$. Also, in the above equation $x(j,l)=\sqrt{3}(l-j/2)$ and $y(j,l)=\frac{3}{2} j$ are the re-defined lattice indices.
The integral in Eq.~\ref{eigenvectorGapless} vanishes as we approach the boundary,  $J_z^2 \to (J_x-J_y)^2$, of the parameter regime defining the gapless phase, since $q^* \to \pi$ here.  
  In the opposite limit of  $J_z^2\to (J_x+J_y)^2$, $q^*\to 0$, and as $J_z$ crosses this condition we move into the gapped phase solution which will be discussed in the next section, Sec.~\ref{Gapped}.

For $J_x=J_y=J_z$, we have $q^*=2\pi/3, \alpha=\sqrt{3},\theta(q)=0$, and thus the result of Pereira et.al.,~\cite{CastroNetoPRL},  $a(x,y)\sim \Re \left\{ (e^{i2\pi x/3\sqrt{3}+ i2\pi y/3})/(y-ix)\right\}$ is recovered. 
A numerically exact zero mode in the gapless phase is shown in Fig.~(\ref{figure2}) for a finite system with periodic boundary conditions. 

\subsection{Gapped phase}
\label{Gapped}

\begin{figure}
\includegraphics[width=4cm,angle=270]{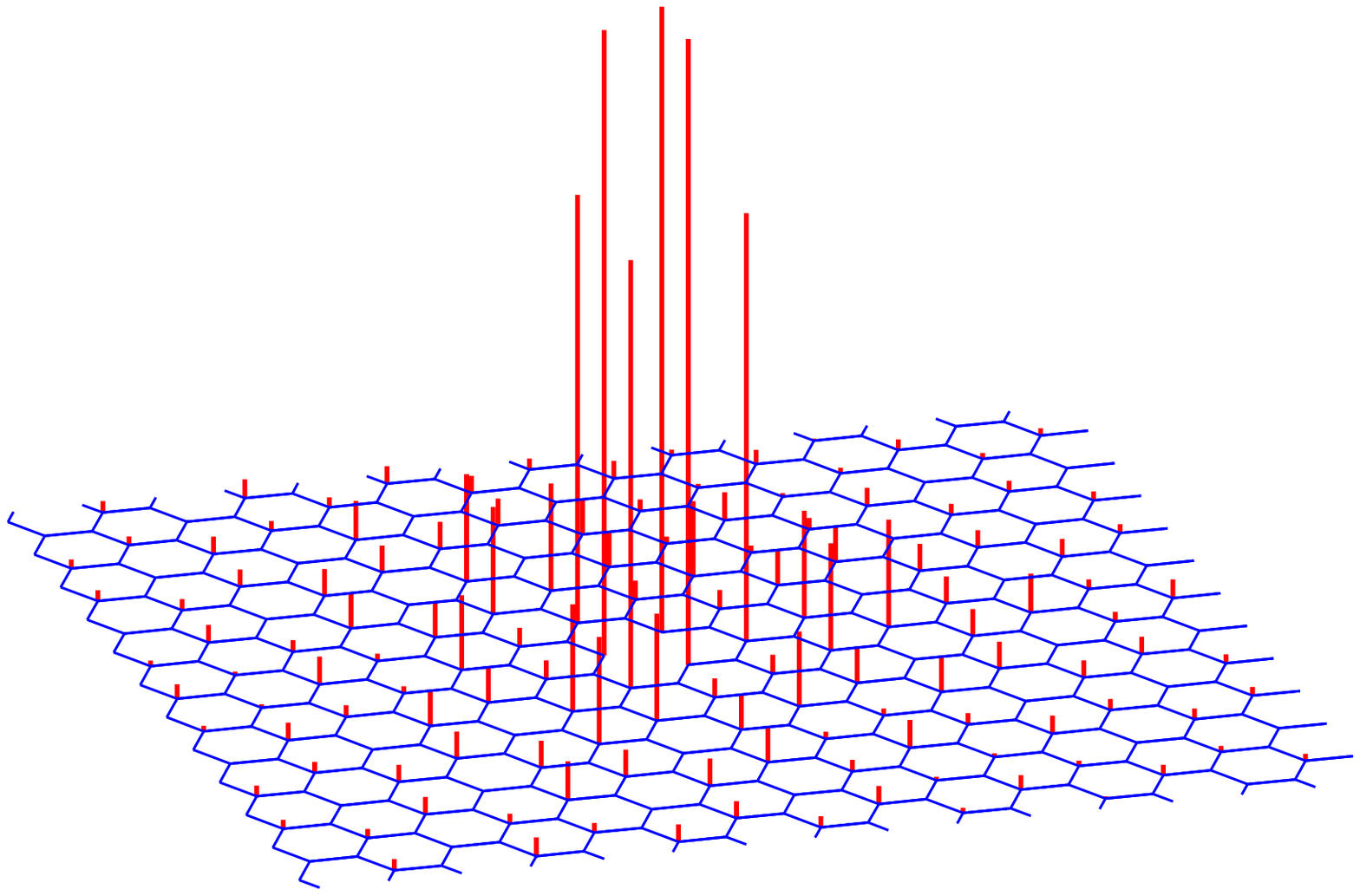}
\includegraphics[width=4cm,angle=270]{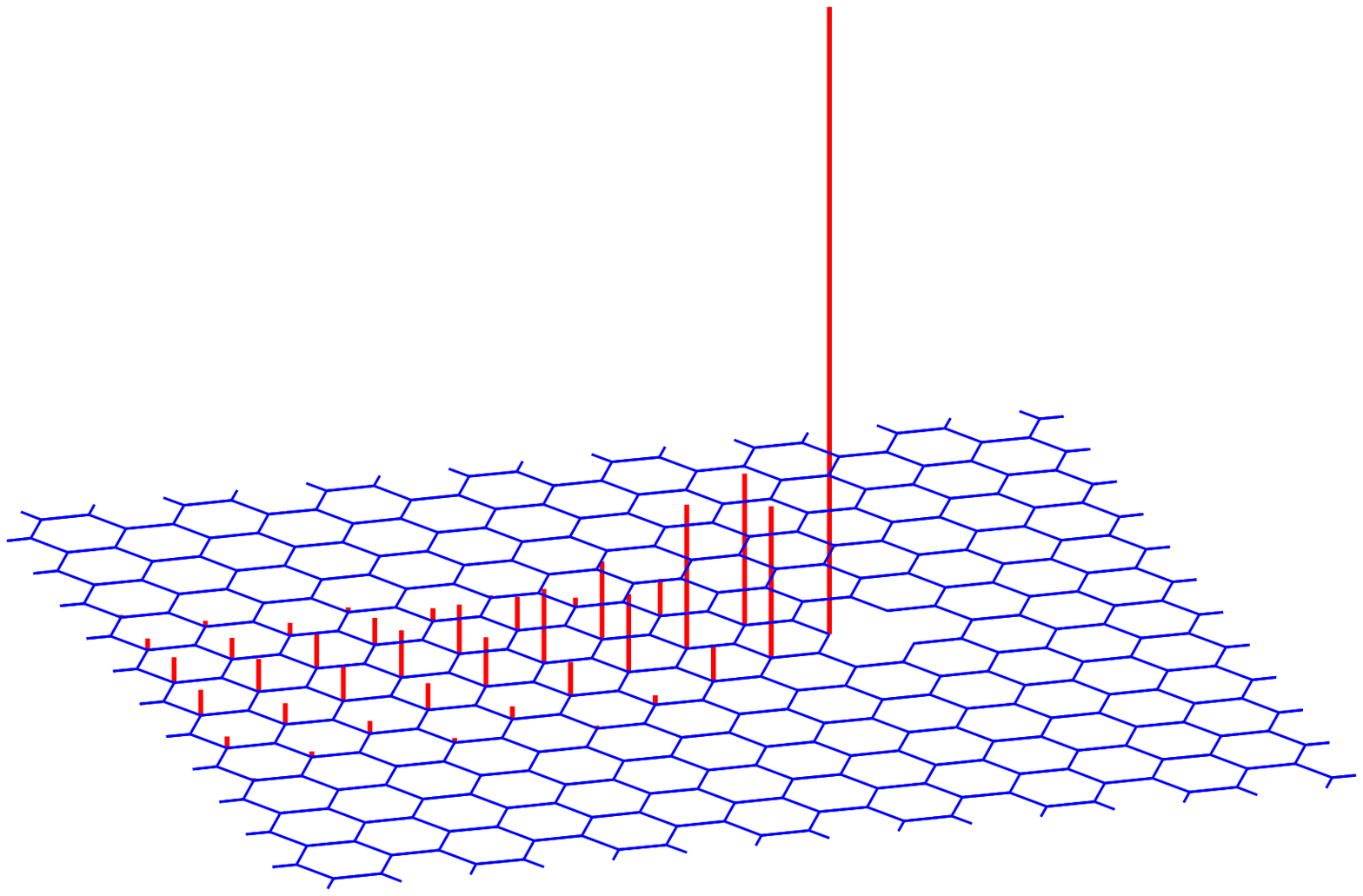}
 \caption{The zero mode intensity $|a_{jl}|^2$ for the gapless case ($J_x=J_y=J_z=1)$ (top) and for the gapped case ($J_x=J_y=1, J_z=2.05$) (bottom) for a system with $N=2500$ unit cells
 and with periodic boundary conditions.}
\label{figure2}
\end{figure}

Now consider the gapped phase. For the sake of concreteness, consider the situation wherein all $J$s are taken to be positive and furthermore satisfy the condition $J_x+J_y<J_z$ . As we have already seen, we have $|f(q)|<1$ for all values of $q$ and hence the solution for $j< 0$ is trivially zero. Then the boundary condition at $j=0$ implies (see Eq.~\ref{eigenvalueEquation}) that
\begin{equation}\label{boundaryconditionGap}
a_{0 l}=0 \quad \forall  \; l\neq 0.
\end{equation}
The solution of Eq.~\ref{eigenvalueEquation} satisfying this boundary condition is 
\begin{eqnarray}
a_{j, l}={(-1)^j}  \; ^j C_{l}\; \left(\frac{J_x}{J_z} \right)^{j-l}  \left(\frac{J_y}{J_z} \right)^{l} a_{0,0},
\end{eqnarray}
for all $j>0, l \in \{0,..,j\}$ and zero everywhere else. Here $^j C_l$ is the Binomial coefficient. We note that the solution is non-zero only in a cone-shaped region extending in the $j>0$ direction. For any $j>0$, 
\begin{eqnarray}
|a_{j,l}| \leq \sum_l |a_{j,l}|=[(J_x+J_y)/J_z]^j a_{0,0}.
\end{eqnarray}
Therefore $|a_{jl}|$ decay exponentially since $(J_x+J_y)/J_z <1$. Note that we have implicitly assumed that the lattice extends infinitely in the $j$ direction. If we have periodic boundary condition in the $j$ direction as well, the tail of this solution can wrap around to the $j<0$ region shown in Fig.~\ref{figLattice}. The zero mode corresponding to the gapped phase, but still close to the transition (at $J_z=J_x+J_y$), is shown in Fig.~(\ref{figure2}). 

We have two other possibilites, namely, $J_y+J_z <J_x$ and $J_z+J_x <J_y$, for the gapped phase. They also give similar results, and are related to the current result by rotation of the lattice by $2\pi/3$ and $4\pi/3$  and cyclic permutation of $J_x,J_y, J_z$.

\subsection{Participation ratio and site-entanglement}
\label{IPR}
\begin{figure}
\includegraphics[width=5cm,angle=270]{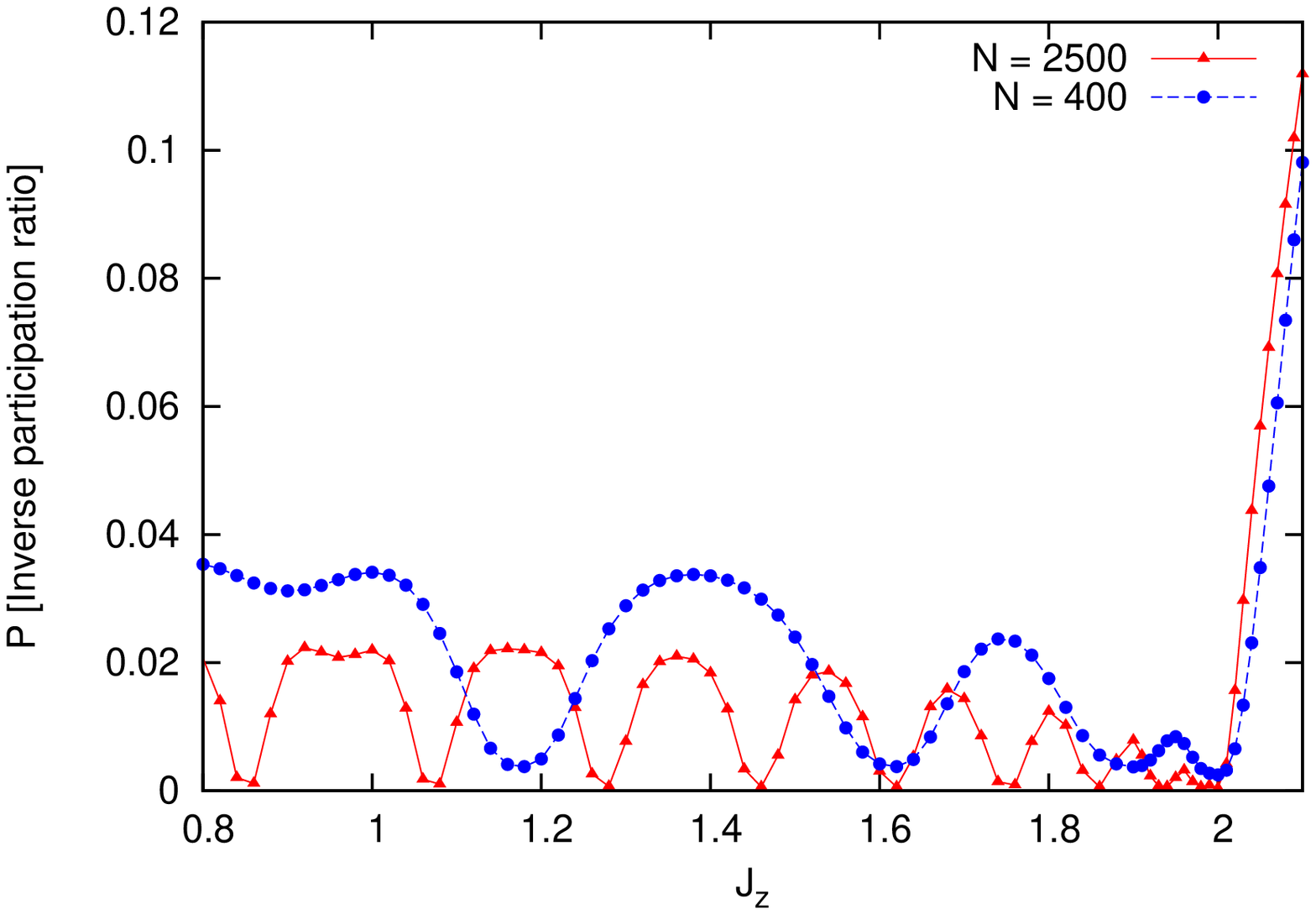}
\includegraphics[width=5cm]{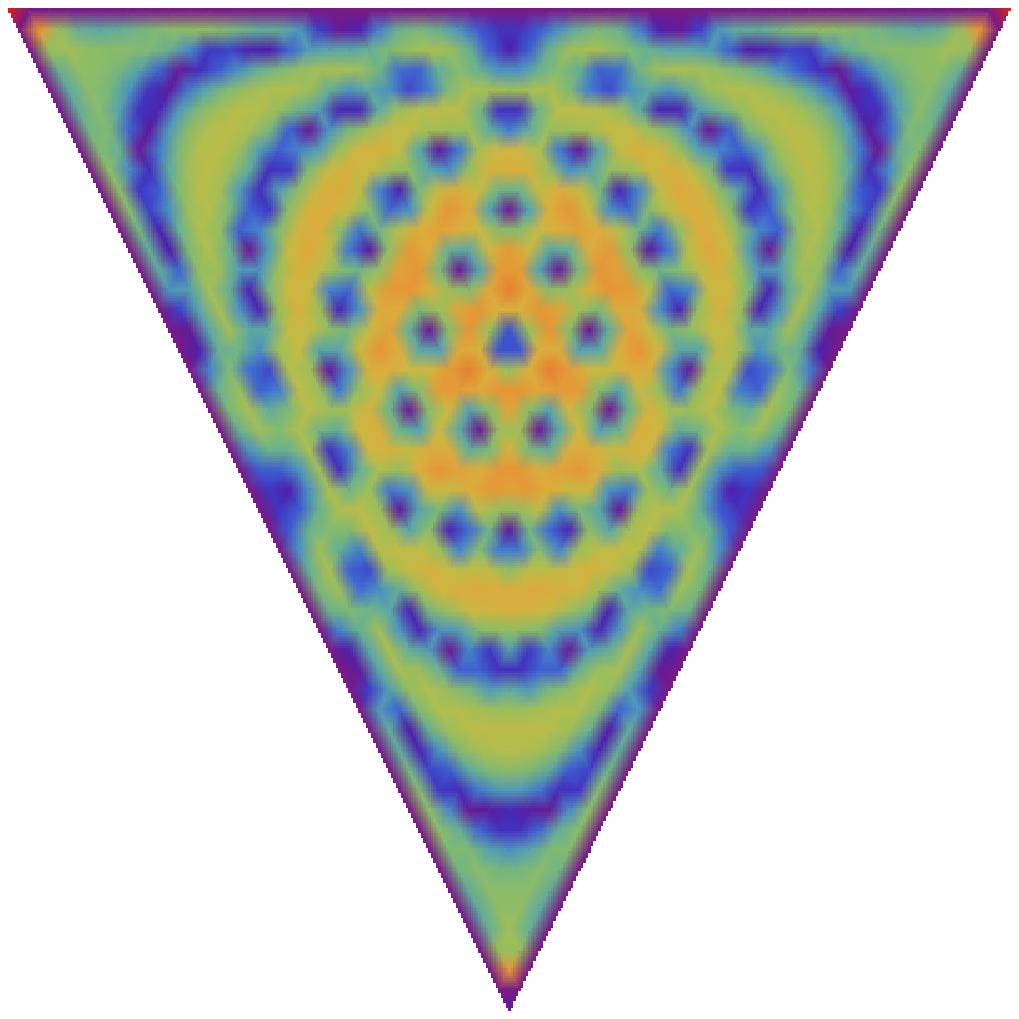}
 \caption{The inverse participation ratio of the zero modes when $J_x=J_y=1$ as a function of $J_z$ (top), the gapless to gapped 
 transition being at $J_z=2$. The IPR as a function of $J_x$, $J_y$ and $J_z$ in the entire gapless phase (bottom) for $N=900$. The darker (blue color) regions have a low IPR or large participation ratio.}
\label{figure3}
\end{figure}

The contrasting nature of the zero modes in the gapped and gapless phases provides motivation for a closer
study. In the gapless  phase there is a ``quasilocalized" zero mode in the terminology of \cite{CastroNeto}, as the amplitude decreases as
$1/r$ from the vacancy. This leads to an anomalous scaling of the inverse participation ratio (IPR) defined as
\begin{eqnarray}
 P=\frac{\sum_{j,l} |a_{jl}|^4}{(\sum_{jl} |a_{jl}|^2)^2}.
\end{eqnarray}
The IPR in the gapless phase would then depend on the size of the system $N$ as $1/\ln(N)^2$~\cite{CastroNeto}, whereas
in the gapped phase the IPR would be independent of the system size reflecting the localized nature of the zero mode. In Fig.~(\ref{figure3}) 
is shown the IPR across a transition to the gapped phase where we can see an increased localization, as indicated by the rapid increase in the IPR
beyond $J_z=2$. 

Quite apart from this dependence, it is interesting to see strong variations of the IPR within the gapless phase as a function
of the parameters $(J_x,J_y,J_z$). In Fig.~(\ref{figure3}), top panel, this is seen in the region $J_z<2$. Also note the 
strong dependence of these oscillations on the system size $N$ in this case.
The variation of the IPR in the entire gapless phase is most neatly captured in the triangle with $J_x+J_y+J_z=1$, 
 with all $J_x$, $J_y$ and $J_z$ being $\le$ $1/2$ \cite{Kitaev}. The IPR of the zero mode for 
parameter values in this triangle corresponding to the gapless phase is shown in Fig.~(\ref{figure3}).  The borders of the triangle corresponding
to an imminent transition to the gapped phase shows a minimum of the IPR, indicating the existence of extended zero modes. 
The dark regions corresponding to very small IPR and large delocalization are arranged in an intriguing manner and require further work for elucidation. The complexity of the figure in terms of the number of such regions with large delocalization increases with the system size $N$.

A trivial  calculation on the gapped side shows that the wavefunction is not only square summable but actually summable:
$\sum_{jl} |a_{j,l}|  \, <\, \infty$.
As the gapless phase boundary is reached this summability is lost. The total site-entanglement present in the one-particle modes is closely related to this sum. Entanglement in the Kitaev model has been recently studied, and 
refers to entanglement in the spins \cite{KitEnt}. However if we are to look at single particle states, we can study entanglement between the sites 
themselves, sites that maybe empty or singly occupied, and the mode is considered to be a superposition of such singly occupied states.

In the context of the Kitaev model the onsite fermions are of Majorana type as opposed to complex ones. Although the site-entanglement measure in the context of Majorana fermions needs further studies, on interpreting the modes as that of a complex fermion hopping problem, site entanglement becomes an especially standard and well studied tool.  
Such entanglement measures have been used previously in many contexts including that of Anderson localization \cite{SudipChakravarty}, wherein
a single site von Neumann entropy has been studied. If however we study the entanglement between a pair of sites, say labelled by $(j,l)$ and $(j',l')$,
the concurrence \cite{Wootters} measure can be used. The concurrence measures entanglement between any two two-state (qubits) systems,
and sites with occupancy 0 or 1 are precisely isomorphic to qubits. If the concurrence is 0, there is no entanglement between the two sites and
if it is 1, they are maximally entangled.
 For one-particle states the concurrence is simply \cite{LakSub} 
\begin{equation}\label{concurrence}
C_{jl.j'l'}=2|a_{j,l} a_{j',l'}|
\end{equation}
and the total concurrence, summed over all pairs of sites is  $C_T$ where 
\begin{equation}\label{totalconcurrence}
C_T= \left( \sum_{j,l} |a_{j,l}| \right)^2 -1 = a_{0,0}^2 \dfrac{J_z^2}{(J_{z}-J_{y}-J_{x})^2}\; -1.
\end{equation}
Thus, at the transition when $J_z=J_x+J_y$ we see a divergence of the total site-site entanglement. 
The inverse participation ratio $P$ is also simply related to site-site entanglement. The sum of the squares
of the concurrence (also called the tangle) across all pairs of sites is related to the IPR. While a closed form 
analytical expression for the IPR seems difficult, as noted above when discussing Fig.~(\ref{figure3}) the IPR is a minimum across the gapless-gapped transition, indicating increased delocalization of the states and large site-site entanglement.

\section{Two vacancies}
\label{twovacancies}
Next we briefly discuss the effects of having a vacancy pair. To do so, let us first consider a sort of index theorem given in Pereira et. al., in the context of graphene \cite{CastroNeto}. 
Generally, this "index"- theorem counts the number of zero modes 
that arise due to presence of vacancy defects in a fermion hopping problem on a bi-partite lattice. 
More specifically, it has been shown that the number of vacancy induced zero modes in such tight-binding type models 
is equal to the difference $|n_B-n_A|$, wherein $n_B(n_A)$ is the number of vacancies on the $A(B)$ 
sub-lattice. Now, it is also known that for instance if $n_B > n_A$, then the zero modes have 
non-zero support only on the  A-sublattice.  The situation is reversed if $n_B<n_A$.  
Thus, according to the above discussion if the vacancy pair is introduced on different sublattices 
then one would assume that the zero-modes interact with each other lifting away from zero. 
While this is indeed true in the gapless case the gapped case comes with an additional wrinkle.   
In other words, in an infinite lattice with open boundary condition, depending on the position of vacancies, there may still be intact zero modes even when the two impurities are placed on different sub-lattices. To see this  let us define $A_{tot}=A+ V_{1,\v{r}_1} +V_{2,\v{r}_2}$.  Then 
\begin{eqnarray}\label{twoimpurity}
\left( A_{tot}\phi_{2,{\v{r}_2}}\right)(\v{r})
&=&2 \phi_a(\v{r}_{12})[ J_y \delta_{\v{r},\v{r}_1} + J_x  \delta_{\v{r},\v{r}_1-(0,1)}\nonumber\\
&&+ J_z \delta_{\v{r},\v{r}_1-(1,1)}] \left(\begin{array}{c}
                                       0\\1
                                      \end{array} \right)
,
\end{eqnarray}
where we have written $\phi\equiv(\phi_a\quad 0)^T$ and $\v{r}_{12}=\v{r}_1-\v{r}_2$. Comparison with the previous section gives $\phi_a(\v{r})=a_{j_r-1,l_r-1}$. For the gapped phase, $\phi_a(\v{r}_{12})$  is zero unless $\v{r}_1$ is within the cone where the zero mode is nonzero. Thus, $\phi_{2,{\v{r}_2}}$ is also a zero mode when $\v{r}_1$ is outside this cone. Note that the fact that the lattice is infinite in the $j$ direction is crucial for this argument. For  periodic boundary condition, the cone could wrap around the torus and the position $\v{r}_1$ will be within the cone.  For other cases including the gapless phase, $\phi_a(\v{r}_{12})$ is nonzero in general and we could represent the effective coefficient matrix in the space spanned by the two zero modes $\phi_{\mu,{\v{r}_{\mu}}}$ as
\begin{eqnarray}
 i\tilde{A}=\left(\begin{array}{cc}
            0& iS\\
-iS& 0
           \end{array}\right),
\end{eqnarray}
where $S=\phi_{1,\v{r}_1}^{T} V_{1,\v{r}_1} \phi_{2,\v{r}_2}$ $=2\phi_a(\v{r}_{12})[J_y \phi_a(0)+J_x \phi_a((0,1))+J_z\phi_a((1,1))]$. Here we have also used the relation $\phi_{1,\v{r}_1}^{T} V_{1,\v{r}_1} \phi_{2,\v{r}_2}=-\phi_{2,\v{r}_2}^{T} V_{2,\v{r}_2} \phi_{1,\v{r}_1}$ which follows from $\phi_{1,\v{r}_1}(\v{r})=\tau^x \phi(\v{r}_1-\v{r})$ and the relation between $V_1$ and $V_2$. The quantity within the square brackets is exactly what is excluded from being zero in Eq.~\ref{eigenvalueEquation}. $S$ is real since the zero modes are real. The eigenvalues of $i\tilde{A}$ are $\pm S$ and the two zero modes lift off from zero eigenvalue and a crude estimate of the new eigenvalues of $iA_{tot}$ is given by $\pm S({\v{r}_{12}})$ that decays with the distance between the two impurities: $S$ decays as powerlaw, asymptotically, with $\v{r}_{12}$ in the gapless phase and exponentially when the A-site impurity is inside the cone defining zero-mode eigenvector for B-site impurity.

\section{Conclusion and Open Problems}
\label{Conclusion}
The role played by vacancies in identifying the gapped and gapless phases has been discussed. In particular it has been shown that a single vacancy in the gapless phase leads to a ``quasi-localized" zero-mode that asymptotically decays as a power-law. In the gapped phase, the zero-mode due to the 
vacancy defect is exponentially localized with-in a cone that emanates from the vacancy. These 
results were obtained analytically by laying recourse to a technique developed in the context of \cite{CastroNeto}, \cite{CastroNetoPRL}. This leads us to conjecture that a single vacancy impurity can act as 
a probe in distinguishing the two-phases of the Kitaev model. These two phases are characterized by very different behaviors of the IPR
as well, and while the transition is characterized by a local minimum of this quantity, it shows for finite lattices, intriguing patterns as a function
of the parameters in the gapless phase. The localization in the gapped phase leads to summable wavefunctions and to a finite
total site-site entanglement as measured by the concurrence. This diverges as the gapless phase is approached in a manner that is
very easy to calculate.

We have also briefly discussed the effect of interacting zero modes. More specifically, 
specializing to the case of two vacancy defects, we have seen that the number of zero-modes in the gapless phase is equal to the difference 
of vacancies in the $A$ and $B$ sub-lattice, in conformity with the ``index"-theorem in \cite{CastroNeto}, \cite{CastroNetoPRL}. We have also argued  that there are situations  in the gapped phase of the Kitaev model, (in the infinite lattice limit with open boundary conditions), wherein the above mentioned ``index"-theorem does not hold.

Now, we turn our attention to some open problems that still remain to be addressed with regards to the 
effect of vacancy disorder on the Kitaev model. As is obvious from this paper, a proper discussion of 
multi-vacancy effect in the Kitaev model is sorely lacking. As a prelude to any such effort one needs to 
generalize the zero-mode counting argument of Pereira et. al. so as to account for the extreme directionality dependence of the zero-mode wave function in the gapped phase. In the limit of multiple 
impurities it is plausible that one can reduce the problem to a system that is governed by an effective free-fermion action with both random mass-term and random $Z_2$ gauge fields. It is an open question whether this is indeed the case. If one could write down such an effective Hamiltonian, in the spirit of \cite{ludwigetal}, \cite{gadewegner}, \cite{gade}, one could analytically investigate the effect of impurities in determining thermodynamic properties like the DoS. It would be interesting to see whether these 
models show Griffiths type behavior exemplified by  a divergent DoS wherein the divergence is controlled by continuously varying exponents that are a function of the disorder concentration, (see 
Ref.~\cite{us_RBIM} where a similar effect was shown to exist in the $\pm J$ random bond Ising model). 
Results that come from such effective action description of disorder effects can also serve to shed light 
on the effect of disorder on spin-liquids in general
  
Some of these issues addressed above could be also studied numerically. More specifically, the 
functional form of the DoS as a function of the disorder concentration and other system parameters 
are being studied by numerical investigations \cite{us_future}.  

As this paper was being written up, we were made aware of a pre-print \cite{willanspre} wherein results 
similar to ours in the context of Kitaev model were obtained. 

\section{Acknowledgements}
We would like to thank G. Baskaran, R. Shankar, and S.R. Hassan for valuable discussions. The authors 
are particularly indebted to F. Evers, and Soumya Bera for a collaboration on a related work, and for their valuable inputs. Finally AL, and RN would like to thank DST, India, for their generous support through the project SR/S2/HEP-012/2009.

\end{document}